\documentclass[12pt]{article}
\usepackage{amsfonts}
\usepackage{amsmath}
\usepackage[onehalfspacing]{setspace}

\input{tcilatex}

\begin{document}

\title{On the degrees of freedom of lattice electrodynamics}
\author{Bo He\thanks{\textit{E-mail address}{\large : he.87@osu.edu}} \ and
F. L. Teixeira \\
{\small ElectroScience Laboratory and Department of Electrical Engineering,}%
\\
{\small \ The Ohio State University, 1320 Kinnear Road, Columbus, OH 43212,
USA}}
\maketitle

\begin{abstract}
Using Euler's formula for a network of polygons for 2D case (or polyhedra
for 3D case), we show that the number of dynamic\textit{\ }degrees of
freedom of the electric field equals the number of dynamic degrees of
freedom of the magnetic field for electrodynamics formulated on a lattice.
Instrumental to this identity is the use (at least implicitly) of a dual
lattice and of a (spatial) geometric discretization scheme based on discrete
differential forms. As a by-product, this analysis also unveils a physical
interpretation for Euler's formula and a geometric interpretation for the
Hodge decomposition.

PACS: 02.60.-x; 02.70.-c; 02.10.Ox; 11.15.Ha

Keywords: Euler's formula; Hodge decomposition; Lattice electrodynamics;
Differential forms\newpage
\end{abstract}

\section{\protect\bigskip Introduction}

There has been continual interest in formulating physical theories on a
discrete lattice \cite{Wilson}\cite{Smit}\cite{Honan}\cite{Chew}. This
"latticization" process not only eliminates the infinities but also provides
a computational route to study the non-perturbative region, which it is
often difficult to handle by purely analytical tools. In this letter, we
shall discuss aspects of a lattice version of classical electrodynamics
based on a geometric discretization \cite{Teixeira}\cite{Teixeira2}.

There are three main approaches to discretize Maxwell equations: finite
differences \cite{Taflove}\cite{Yee}, finite elements \cite{Bossavit} and
finite volumes \cite{Taflove}. However, naive implementations of these
finite methods to discretize Maxwell equations on irregular lattices are
known to cause problems, such as spurious modes \cite{Sun} and late-time
unconditional instabilities \cite{Schuhmann}, that destroy the solutions.
This is often a consequence of the failure to capture some essential physics
of the continuum theory.

By applying some tools of algebraic topology and a discrete analog of
differential forms, classical electrodynamics can be constructed from first
principles on a lattice \cite{Teixeira}. The purpose of this paper is to
show that via such geometric discretization, \ the equality (necessary, but
often not trivially observed in common discretization schemes) between the
number of \textit{\ dynamic} degrees of freedom ($DoFs$) of the electric
field and the number of \textit{dynamic} $DoFs$ of the magnetic field is a
simple consequence\ of Euler's formula for a network of polygons for 2+1
case or volume polyhedra in the 3+1 case. As a by-product, this analysis
also unveils a \textit{physical} interpretation for Euler's formula and a 
\textit{geometric} interpretation for the Hodge decomposition.

\section{Lattice electromagnetic theory}

For simplicity, we consider first a $TE$ field in 2+1 dimensions. The
extension to 3+1 dimensions is considered in Section 3.2.

Maxwell equations in the Fourier domain are written in terms of differential
forms as \cite{Teixeira} 
\begin{eqnarray}
dE &=&i\omega B  \notag \\
\text{ }dB &=&0  \notag \\
\text{ }dH &=&-i\omega D+J  \notag \\
\text{ }dD &=&Q  \label{Maxwell}
\end{eqnarray}%
In the 2+1 case, $H$ is a 0-form, $D$ , $J$ and $E$ are 1-forms, and $B$ and 
$Q$ are 2-forms. The operator $d$ is the exterior derivative. From the
identity $d^{2}=0$, the electric current density $J$ and $\ $the charge
density $Q$ satisfy the continuity equation 
\begin{equation}
dJ=i\omega Q
\end{equation}%
Constitutive equations, which include all metric information in this
framework, are written in terms of Hodge star operators (that fix an
isomorphism between $p$ forms and $2-p$ forms in the 2+1 case) \cite%
{Teixeira}

\begin{equation}
D=\star_{e}E\text{ , }B=\star_{h}H  \label{cons}
\end{equation}

In numerical simulations, because of limited memory, the computational
domain is taken as a closed domain $\Omega $ with boundary $\partial \Omega
. $ In the lattice, the boundary $\partial \Omega $ is approximated by a set
of linked edges $\partial \widehat{\Omega }$ (see Fig.1).

The latticization corresponds to tiling $\Omega $ with a finite number $N_{F}
$ of polygons $\Xi _{m}$, $m=1,..,N_{F}$, of arbitrary shape(see Fig.2) 
\begin{equation}
\Omega \simeq \widehat{\Omega }=\underset{m=1}{\overset{N_{F}}{\oplus }}\Xi
_{m}  \label{primallattice}
\end{equation}%
We require the tiling to be \textit{conformal} i.e., two polygons are either
connected by one single edge or are not connected at all (see Fig.3). These
polygons should also be oriented, forming a \textit{cell-complex} (see
Fig.4) \cite{Teixeira}\cite{Darling}.We denote such oriented tiling (Fig.4)
the \textit{primal} lattice. From the primal lattice, one can construct a 
\textit{dual }lattice by connecting interior points of each adjacent polygon 
\footnote{%
Different \textit{geometrical} constructions can be used to define the dual
lattice. We will not dwell here into such discussion since our conclusions
depend only on \textit{topological} properties (connectivity) of the lattice.%
}. The dual lattice inherits an orientation from the primal lattice.

Now we consider casting Maxwell equations on a lattice using the natural
latticization provided by casting differential forms of various degrees $p$
in Eq. (\ref{Maxwell}) as dual elements (\textit{cochains})\textit{\ }to $p$
dimensional geometric constituents of the lattice, i.e., \textit{cells}:
nodes, edges and faces \cite{Teixeira}. In the primal lattice, we associate
the electrostatic potential $\phi $ ($0$-form ) with primal nodes ($0$%
-cells), the electric field intensity $E$ ($1$-form) with primal edges ($1$%
-cells) and the magnetic flux density $B$ ($2$-form) with primal cells ($2$%
-cells). In the dual lattice, we associate the magnetic field intensity $H$ (%
$0$-form ) with dual nodes ($0$-cells), \ the electric flux density $D$ ($1$%
-form), the electric current density $J$ ($1$-form) with dual edges ($1$%
-cells), and the charge density $Q$ ($2$-form) with dual cells ($2$-cells).
This is illustrated in Fig.5.

The exterior derivative $d$ can be discretized via its adjoint operator, the
boundary operator, $\partial $, by applying the generalized Stoke's theorem
on each $p$-cell of the cell-complex%
\begin{equation}
\left\langle \gamma ^{p+1},d\alpha ^{P}\right\rangle =\left\langle \partial
\gamma ^{p+1},\alpha ^{P}\right\rangle \text{ ,}
\end{equation}%
where $\gamma ^{p+1}$ is a p+1 dimensional cell and $\alpha ^{P}\in
F^{p}\left( \Omega \right) $, with $F^{p}\left( \Omega \right) $ being the
space of differential forms (cochains in the discrete setting) of degree $p$
on the domain $\Omega $. We denote an ordered sequence of the above pairing
of cochains with each of the cells by block letters $\mathbb{E}$, $\mathbb{B}
$, $\mathbb{H},\mathbb{D}$, $\mathbb{J}$, $\mathbb{Q}$ in what follows.
These are the $DoFs$ of the lattice theory. In terms of these $DoFs$, the
lattice analog of Maxwell's equations is written as \cite{Teixeira} \cite%
{Hiptmair2003} 
\begin{eqnarray}
C\mathbb{E} &\mathbb{=}&i\omega \mathbb{B}\text{ }  \label{network} \\
\text{ }S\mathbb{B} &\mathbb{=}&0 \\
\text{ }\widetilde{C}\mathbb{H} &\mathbb{=}&\mathbb{-}i\omega \mathbb{D+J} \\
\text{ }\widetilde{S}\mathbb{D} &\mathbb{=}&\mathbb{Q}
\end{eqnarray}%
In the above, $C$,$\widetilde{C},S,\widetilde{S}$ are the \textit{incidence
matrices} \cite{Teixeira}, obtained by applying the boundary operator $%
\partial $ on each $p$-cell (discrete version of the exterior derivative $d$%
). \ Because the exterior derivative $d$ is a purely topological operator
(metric-free), the incidence matrices represent pure combinatorial
relations, whose entries assume only $\left\{ -1,0,1\right\} $ values.

The lattice version of the Hodge isomorphism can be, in general, written as
follows 
\begin{equation}
\mathbb{D}=\left[ \star _{e}\right] \mathbb{E}\text{, }\mathbb{B}=\left[
\star _{\mu }\right] \mathbb{H}
\end{equation}%
where both $\left[ \star _{e}\right] $ and $\left[ \star _{\mu }\right] $
are square invertible matrices. We will not discuss here how to construct
the Hodge matrices $\left[ \star _{e}\right] $ and $\left[ \star _{\mu }%
\right] $. That the metric dependent Hodge matrices are the square
invertible matrices is the only assumption needed here.

In many problems of interest, one is usually interested only in the dynamic
behavior of the field, which is determined by the dynamic $DoFs$ of lattice
theory. Here we show that the number of dynamic $DoFs$ of the electric
field, $DoF^{d}\left( E\right) $, equals to the number of dynamic $DoFs$ of
the magnetic flux, $DoF^{d}\left( B\right) $, in the discretization above.
Furthermore, from the isomorphisms between $E$ and $D$, and between $H$ and $%
B$ (from the Hodge maps), this also implies%
\begin{equation}
DoF^{d}\left( E\right) =DoF^{d}\left( B\right) =DoF^{d}\left( D\right)
=DoF^{d}\left( H\right)   \label{dmcp}
\end{equation}%
where the superscript $d$ stands for dynamic. Moreover, for 2+1 $TE$ field,
this number equals to the total number of polygons used for tiling the
domain $\Omega $ minus one ($N_{F}-1$).

\section{Hodge decomposition}

The Hodge decomposition can be written in general as 
\begin{equation}
F^{p}\left( \Omega \right) =dF^{p-1}\left( \Omega \right) \oplus \delta
F^{p+1}\left( \Omega \right) \oplus \chi ^{p}\left( \Omega \right)
\label{generalHelholtz}
\end{equation}%
where $\chi ^{p}\left( \Omega \right) $ is the\textit{\ finite} dimensional
space of harmonic forms, and $\delta $ is the codifferential operator,
Hilbert adjoint of $d$ \cite{westenholz}. Applying (\ref{generalHelholtz})
to the electric field intensity $E$, we obtain

\begin{equation}
E=d\phi +\delta A+\chi  \label{Helmholtz}
\end{equation}%
where$\ \phi $ is a $0$-form and $A$ is a $2$-form. In Eq. (\ref{Helmholtz}) 
$d\phi $ represents the static field and $\delta A$ represents the dynamic
field, and $\chi $ represents \ the harmonic field component.

\subsection{2+1 theory in a contractible domain}

If domain $\Omega $ is contractible, $\chi $ is identically zero and the
Hodge decomposition can be simplified to

\begin{equation}
E=d\phi +\delta A  \label{Hodge}
\end{equation}%
\ In the present lattice model, the number of $DoFs$ for the zero eigenspace 
$\left( \omega =0\right) $ equals the number of internal nodes of the primal
lattice \cite{Sun} \cite{Arnold} \cite{peterson}. This is because the $DoFs$
of the potential $\phi $, which is a $0$-form, is associated to nodes.

Now we show the identity (\ref{dmcp}). Recall the Euler's formula for a
general network of polygons without holes (Fig.2) 
\begin{equation}
N_{V}-N_{E}=1-N_{F}  \label{Euler's formula1}
\end{equation}%
Here, $N_{V}$ is the number of vertices (nodes), $N_{E}$ the number of
edges, and $N_{F}$ the number of faces (cells). For any $\partial \widehat{%
\Omega }$, it is easy to verify that 
\begin{equation}
N_{V}^{b}-N_{E}^{b}=0  \label{close-1-chain}
\end{equation}%
where $N_{V}^{b}$ is the number of vertices on the boundary and $N_{E}^{b}$
the number of edges on the boundary (the superscript $b$ stands for
boundary). Note that cochains on $\partial \widehat{\Omega }$ are not
associated to $DoFs$, since they are fixed from the boundary conditions.
Using the Hodge decomposition (\ref{Helmholtz}), the number of dynamic ($%
\omega \neq 0$ ) $DoFs$ of the \ electric field, corresponding to $\delta A,$
is given by 
\begin{eqnarray}
DoF^{d}\left( E\right)  &=&N_{E}^{in}-N_{V}^{in}  \notag \\
&=&\left( N_{E}^{{}}-N_{E}^{b}\right) -\left( N_{V}^{{}}-N_{V}^{b}\right)  
\notag \\
&=&N_{E}^{{}}-N_{V}  \label{Efreedom}
\end{eqnarray}%
where the superscript $in$ stands for internal. Since $E$ is given along the
boundary, then, for $\omega \neq 0,$ $\int_{\widehat{\Omega }}B$ is fixed by 
$\ $%
\begin{equation}
i\omega \int_{\widehat{\Omega }}B=\int_{\partial \widehat{\Omega }}E
\label{constrain}
\end{equation}%
This corresponds to \textit{one} constraint on $B$. Subtracting one degree
of freedom from the constraint (\ref{constrain}), the number dynamic $DoFs$
of the magnetic flux $B$ is%
\begin{equation}
DoF^{d}\left( B\right) =N_{F}-1  \label{Hfreedom}
\end{equation}%
From Euler's formula (\ref{Euler's formula1}), we then have the identity 
\begin{equation}
DoF^{d}\left( E\right) =DoF^{d}\left( B\right) 
\end{equation}%
Furthermore, thanks to the Hodge isomorphism, the identity (\ref{dmcp})
follows directly.

\subsection{3+1 theory in a contractible domain}

The source free Maxwell equations in 3+1 dimensions read as%
\begin{eqnarray}
dE &=&i\omega B, \\
\text{ }dB &=&0,  \label{constraint} \\
\text{ }dH &=&-i\omega D, \\
\text{ }dD &=&0
\end{eqnarray}%
where now $H$ and $E$ are 1-forms, and $D$ and $B$ \ are 2-forms. The
spatial domain $\Omega $ is again (approximately) tiled by a \textit{set of
polyhedra} $\widehat{\Omega }$ \ and the boundary $\partial \Omega $ is by a%
\textit{\ polyhedron }$\partial \widehat{\Omega }$ . Using Euler's formula
for $\widehat{\Omega }$, we have

\bigskip 
\begin{equation}
N_{V}^{{}}-N_{E}^{{}}=1-N_{F}^{{}}+N_{P}^{{}}  \label{Euler31}
\end{equation}%
and Euler's formula for the boundary polyhedron $\partial \widehat{\Omega }$ 
\begin{equation}
N_{V}^{b}-N_{E}^{b}=2-N_{F}^{b}  \label{Euler3b}
\end{equation}%
\bigskip where $N_{P}$ is now the number of polyhedra. Combining Eq. (\ref%
{Euler31})\ \ and (\ref{Euler3b}), we obtain%
\begin{equation}
\left( N_{E}^{{}}-N_{E}^{b}\right) -\left( N_{V}^{{}}-N_{V}^{b}\right)
=\left( N_{F}^{{}}-N_{F}^{b}\right) -\left( N_{P}^{{}}-1\right) 
\label{Euler3}
\end{equation}%
Using the Hodge decomposition (\ref{Hodge}), the number of dynamic $DoFs$ of
the \ electric field (corresponding to $\delta A$ ) is

\begin{eqnarray}
DoF^{d}\left( E\right)  &=&N_{E}^{in}-N_{V}^{in}  \notag \\
&=&\left( N_{E}^{{}}-N_{E}^{b}\right) -\left( N_{V}^{{}}-N_{V}^{b}\right) 
\label{Edof}
\end{eqnarray}%
Each polyhedron produces one constraint for the magnetic flux $B$ from Eq.(%
\ref{constraint}). Furthermore, this set of constraints span the condition
at the boundary $\partial \widehat{\Omega }$. The total number of the
constrains for $B$ is therefore $\left( N_{P}^{{}}-1\right) .$ Consequently,
the number of $DoFs$ for the magnetic flux $B$ is%
\begin{eqnarray}
DoF^{d}\left( B\right)  &=&N_{F}^{in}-\left( N_{P}^{{}}-1\right)   \notag \\
&=&\left( N_{F}^{{}}-N_{F}^{b}\right) -\left( N_{P}^{{}}-1\right) 
\label{Bdof}
\end{eqnarray}%
Identity (\ref{dmcp}) then follows from Eq. (\ref{Euler3}), (\ref{Edof}) and
(\ref{Bdof}).

\subsection{2+1 theory in a non-contractible domain}

Now consider a non-contractible two-dimensional domain $\Omega $ with a
finite number $g$ of holes (genus). This is illustrated in Fig. 6 for $g=1.$
Along the boundary of each hole, the electric field $E$ is constrained by 
\begin{equation}
\int E=M  \label{constrain2}
\end{equation}%
where the magnetic current density \footnote{%
In physical terms, \ the magnetic current density $M$ is identified with the
"displacement magnetic current density" $i\omega B$, which is given for some
cases. In some other cases, $M$ comes also from the \textit{equivalent}
magnetic current density by the surface equivalence theorem \cite{Balanis}.
It should be emphasized that the \textit{equivalent} magnetic current
results from an impressed electric field $E$, not from the movement of any
"magnetic charge".} $M$ (passing through the hole) is a known quantity. The
equation (\ref{constrain2}) accounts for the possible existence of the
harmonic forms $\chi $ on $\Omega $. In particular, the number of holes $g$
is equal to the dimension of the space of harmonic forms $\chi $ and gives
the number of independent constraint equations (\ref{constrain2}).
Subtracting $g$ from Eq. (\ref{Efreedom}), the number of dynamic $DoFs$ of
the \ electric field in this case becomes 
\begin{eqnarray}
DoF^{d}\left( E\right)  &=&N_{E}^{in}-N_{V}^{in}-g  \notag \\
&=&N_{E}^{{}}-N_{V}-g  \label{Efreedom2}
\end{eqnarray}%
whereas the number of $DoFs$ of the magnetic flux $DoF^{d}\left( B\right) $
remains $N_{F}-1.$

Since Euler's formula for a network of polygons with $g$ holes is%
\begin{equation}
N_{V}-N_{E}=\left( 1-g\right) -N_{F}  \label{Eulerformula2}
\end{equation}%
we have that from Eq. (\ref{Hfreedom}), (\ref{Efreedom2}) and (\ref%
{Eulerformula2}), the identity (\ref{dmcp}) is again satisfied.

\subsection{\protect\bigskip Euler's formula and Hodge decomposition}

From the above considerations, \ we can trace the following correspondence
in the 2+1 case

\begin{equation}
\begin{array}[t]{ccccccc}
N_{E} & = & N_{V} & + & \left( N_{F}-1\right) & + & g \\ 
\updownarrow &  & \updownarrow &  & \updownarrow &  & \updownarrow \\ 
E & = & d\phi & + & \delta A & + & \chi%
\end{array}
\label{relation}
\end{equation}%
The number of edges $N_{E}$ corresponds to the dimension of the space of
(discrete) electric field intensity $E$ ($1$-forms), which is the sum of the
number of nodes $N_{V}$ (dimension of the space of discrete $0$-forms $\phi $%
), the number of faces $\left( N_{F}-1\right) $ (dimension of the space of
discrete $2$-form $A$) and the number of holes $g$ (dimension of the space
of harmonic form $\chi $). These correspondences attach a physical meaning
to Euler's formula and a geometric interpretation to the Hodge
decomposition. We note that the identity (\ref{relation}) can be viewed as%
\begin{equation}
\begin{array}[t]{ccccccc}
N_{E}^{in} & = & N_{V}^{in} & + & \left( N_{F}-1\right) & + & g \\ 
\updownarrow &  & \updownarrow &  & \updownarrow &  & \updownarrow \\ 
E & = & d\phi & + & \delta A & + & \chi%
\end{array}%
\end{equation}%
since only the internal edges and nodes describe the degrees of freedom. We
can simply drop the superscript $in$ because the identity (\ref%
{close-1-chain}). For the 3+1 case, a similar correspondence could also be
drawn.

\section{Concluding remarks}

Based on a geometric discretization\footnote{%
One key feature of this scheme is the use of a dual lattice and of a
geometric discretization scheme based on differential forms, also proposed
in different contexts in \cite{Adams}\cite{Sen}\cite{Rabin}.}, we have shown
that Euler's formula matches the algebraic properties of the discrete
Helmholtz decomposition in an exact way. Furthermore, we have showed that
the number of dynamic $DoFs$ for the electric field equals the number of
dynamic $DoFs$ for the magnetic field on such lattices\footnote{%
For the case of high order $1$-forms \cite{Ren}, the $DoFs$ of $1$-forms
could associate with the faces and volumes. \ However, the dimension of the
range space of 1-forms (e.g. $E$) equals the dimension of 2-forms (e.g. $B$)
if such 2-forms have zero range space due to $dB=0$, so the identity (\ref%
{dmcp}) still holds.
\par
{}}.

Regarding the time discretization, we also remark that simplectic
integrators, originally developed for Hamiltonian systems \cite{Sanz}, \ can
provide a time discretization that respects the simplectic structure \cite%
{Kole1}. However, it is not trivial to formulate the Maxwell's equations on
a lattice as the canonical equations of the Hamiltonian, because
electrodynamics can be thought of\ as a\textit{\ constrained} dynamic
system. A Hamiltonian requires that the canonical pair ($E,H$) should have
the same number of degrees of freedom. Identity (\ref{dmcp}) suggests that
it is indeed possible to formulate electrodynamics on a lattice as the
canonical equations of the Hamiltonian.

\bigskip

\textbf{Acknowledgements}

This work has been supported in part by NSF under grant ECS-0347502.

\newpage

\begin{center}
\textbf{Figure captions}
\end{center}

Fig.1. The curved boundary $\partial \Omega $ is approximated by a set of
linked edges $\partial \widehat{\Omega }$.

\bigskip

Fig.2. Tiling the computation region with polygons.

\bigskip

Fig.3 (a) conformal tiling (cell complex); (b) non-conformal tiling.

\bigskip

Fig. 4. Oriented polygons.

\bigskip

Fig. 5. Solid lines represent the primal lattice. Primal nodes (vertices)
are paired with $\phi $ (e.g., node $1$), primal edges with $E$ (e.g., edge $%
15)$ and primal cells with $B$ (e.g., cell $12345).$ Dashed lines represent
the dual lattice. Dual nodes are paired with $H$ (e.g., node $4^{\prime }$),
dual edges with $(D,J)$ (e.g., edge $3^{\prime }4^{\prime }$) and dual cells
with $Q$ (e.g., cell $1^{\prime }2^{\prime }3^{\prime }4^{\prime }5^{\prime }
$).

\bigskip

Fig. 6. 2+1 theory in a non-contractible domain (network of polygons with a
hole, illustrated by a triangle $123$).

\end{document}